\documentclass[aps,pra,twocolumn,showpacs]{revtex4}
\usepackage[dvips]{graphicx}
\usepackage[normalem]{ulem}
\usepackage{graphicx}
\usepackage{dcolumn}
\usepackage{bm}
\input{epsf}
\usepackage{color}

\newcommand{\be}{\begin{equation}}
\newcommand{\ee}{\end{equation}}
\newcommand{\beq}{\begin{eqnarray}}
\newcommand{\enq}{\end{eqnarray}}
\newcommand{\ua}{\uparrow}
\newcommand{\da}{\downarrow}
\newcommand{\uden}{n_\uparrow}
\newcommand{\dden}{n_\downarrow}

\newcommand{\eup}{E_{F \uparrow}}
\newcommand{\om}{\omega}

\newcommand{\rms}{\rm S}
\newcommand{\rmn}{\rm N}
\newcommand{\eps}{\epsilon}
\newcommand{\mup}{m_{\uparrow}}
\newcommand{\mdown}{m_{\downarrow}}

\newcommand{\xc}{x_{\rm c}}
\newcommand{\etc}{\eta_{\rm c}}
\newcommand{\ka}{\kappa}
\newcommand{\mua}{\mu_{\uparrow}}
\newcommand{\mda}{\mu_{\downarrow}}

\newcommand{\as}{\alpha_{\sigma}}
\newcommand{\au}{\alpha_{\uparrow}}
\newcommand{\ad}{\alpha_{\downarrow}}

\begin{document}

\title{Chandrasekhar-Clogston limit and phase separation in Fermi mixtures at Unitarity}  

\author{I. Bausmerth$^{1}$}\email{ingrid@science.unitn.it}
\author{A. Recati$^{1,2}$}
\author{S. Stringari$^{1}$}
\affiliation{$^{1}$Dipartimento di Fisica, Universit\`a di Trento and CNR-INFM BEC,
I-38050 Povo, Italy\\
$^{2}$Physik Department, Technische Universit\"at M\"unchen, D-85747 Garching, Germany} 

\begin{abstract}
Using many-body results available from diagrammatic and {\it ab initio} Monte Carlo calculations we analyze the phase diagram $\mu=(\mua+\mda)/2$ versus  $h=(\mua-\mda)/2$ of a unitary Fermi gas at zero temperature with population imbalance and unequal masses. We identify different regions where the gas is superfluid, partially polarized or fully polarized and determine the corresponding coexistence conditions. The asymmetry in the phase diagram, caused by the mass imbalance, and its effect on the Chandrasekhar-Clogston limit for the critical polarization are explicitly discussed. The equation of state of the superfluid and normal  phases is employed, within the local density approximation, to predict phase separated configurations in the presence of harmonic trapping potentials.

\end{abstract}


\maketitle

\section{Introduction}

The study of polarization effects in Fermi superfluids has been the object of intense experimental and theoretical work (for recent reviews on the subject see, e.g., \cite{ketterle,rmp}) in ultracold atomic gases in the last few years. Crucial goals of these studies are the identification of quantum phases and the determination of the Chandrasekhar-Clogston limit of critical polarization above which the system is no longer superfluid.

The recent observation of heteronuclear Feshbach resonances in ultracold mixtures of two fermionic species \cite{wille} as well as the realization of a degenerate two-species Fermi-Fermi mixture \cite{haensch} has opened new stimulating perspectives in the field of  Fermi superfluids built with atomic species of different masses. 

The  phase diagram of  Fermi mixtures with unequal masses and the corresponding polarization effects, including the possible occurrence of the Fulde-Ferrell-Larkin-Ovchinnikov (FFLO) phase \cite{iskin}, have already been the object of theoretical predictions based on BCS mean-field theory \cite{rupak,yip,duan,lama,melo}. This theory is known to give reasonable predictions at unitarity in the case of unpolarized configurations (see, for example, \cite{rmp}). However, it fails to give quantitatively correct results in the imbalanced case and to predict the Chandrasekhar-Clogston limit of critical polarization. The failure of the BCS mean-field is mainly due to the fact that it ignores the role of interactions in the normal phase which are now understood to play a crucial role at unitarity \cite{normal,alessio}.

\begin{figure}[htb]
	\centering
		\includegraphics[height=5cm]{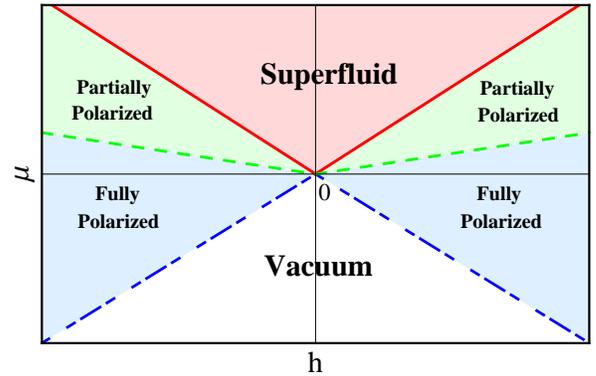}
	\caption{(Color online) In the case of equal masses the $\mu$-$h$ phase diagram is symmetric with respect to zero effective magnetiv field $h$. Shown are the superfluid ($\rm S$, red), the partially polarized ($\rm PP$, green) and fully polarized ($\rm FP$, blue) phases.}
	\label{fig:Apd}
\end{figure}

The main goal of this paper is to use the present knowledge of the equation of state of Fermi mixtures with unequal masses to give quantitative predictions for the phase separation between the normal and superfluid components. Our analysis is based on the study of the zero temperature $\mu$-$h$ phase diagram as shown in Fig. \ref{fig:Apd} of the uniform two component gas, where $\mu=(\mua+\mda)/2$  is the chemical potential and $h=(\mua-\mda)/2$ is an effective magnetic field.
The phase diagram at unitarity is determined thanks to the knowledge of the equation of state available from diagrammatic techniques applied to highly polarized configurations and from Monte Carlo simulations.
The phase diagram is then used, in the local density approximation (LDA), to calculate the density profiles of the two Fermi components in the presence of harmonic trapping. 

We begin in Sec. \ref{normalstate} by reviewing the general theory of the normal state for equal masses, and discuss its extension to the unequal mass case. Then in Sec. \ref{pd} we introduce the bulk phase diagram and discuss its properties in dependence on the mass ratio $\ka$. In Sec. \ref{trap} we use the phase diagram in local density approximation to calculate the density profiles, focusing on three particular configurations. Finally in Sec. \ref{con} we draw our conclusions.

\section{Normal State of a Fermi Gas with unequal masses}
\label{normalstate}

The equation of state of the normal phase in the unitary limit of infinite scattering length and at zero temperature was first derived by Lobo \textit{et al.} \cite{normal} in the case of equal masses using the concept of quasiparticles.
As a function of the concentration $x=\dden/\uden$ and for $x\ll1$ it is given by
\beq
\frac{E(x)}{N_\ua}&=&\frac{3}{5}\eup\left(1-Ax+\frac{m}{m^*}x^{5/3}+B x^2\right), 
\label{eq:eoslobo}
\enq
where $N_\ua$ is the total number of spin-$\ua$ atoms and $\eup=\hbar^2/2m (6 \pi^2 \uden)^{2/3}$ the Fermi energy of the spin-$\ua$ gas.
In Eq.(\ref{eq:eoslobo}) it is assumed that adding a few spin-$\da$ particles to a noninteracting spin-$\ua$ sea, the $\da$-atoms form a Fermi gas of quasiparticles with an effective mass $m^*$.

The first term  in Eq.(\ref{eq:eoslobo}) corresponds to the energy per particle of the noninteracting gas, while the term linear in $x$ gives the binding energy of the spin-$\da$ particles to the spin-$\ua$ sea. 
The interaction between $\ua$ and $\da$ particles is accounted for by the parameter $A=0.99(1)$ \cite{sebastiano}, which is proportional to the ratio $|\mu_{\da}|/\mu_{\ua}$.
The Fermi gas of quasiparticles with an effective mass $m^{\ast}$ contributes to the total energy in Eq.(\ref{eq:eoslobo}) by the quantum pressure term proportional to $x^{5/3}$, and in the case of equal masses $m^*/m=1.09(2)$ \cite{sebastiano}.
Eventually, the term proportional to $x^2$ can be interpreted as an interaction between quasiparticles, and its value $B=0.14$ has been determined fitting the expression (\ref{eq:eoslobo}) to the Monte Carlo results for the equation of state as a function of the concentration \cite{sebastiano}.
Although based on a small $x$ expansion, Eq.(\ref{eq:eoslobo}) turns out to account for the $x$-dependence of the equation of state also for values of $x$ close to 1.

For $\mup\neq\mdown$ the values of the parameters $A$ and $m^*$ depend on the mass ratio in a non-trivial way and have been calculated in \cite{Combescot} and \cite{Comb2} as functions of the mass ratio $m_{\da}/m_{\ua}=\ka$ using diagrammatic many-body techniques. The parameter $A$ is an increasing function of the mass ratio $\ka$ going to infinity for $\ka\rightarrow 0$ and reaching the asymptotic value $A\sim0.45$ for $\ka\rightarrow\infty$.
On the other hand at unitarity the effective mass, which we will denote in the rest of the paper as ${m^*}={m_{\da}}F(\ka)$, shows a weak dependence on the mass ratio.

The quasiparticle interaction $B$ has up to now only been determined for equal masses $\mdown=\mup$. 
We can find a first estimation for $B(\ka)$ in the following way.
In the unpolarized case ($x=1$) the energy of the normal state as a function of the mass ratio $\ka$ has been calculated using Monte Carlo methods \cite{Carlsonxi,Stefanoxi}, resulting in the expression 
\be
\frac{E_{\rmn}(\ka)}{N_{\ua}}=\xi_{\rmn}(\ka)\frac{3}{5}\frac{\hbar^2}{4 m_{\rm \ka}} (6 \pi^2 \uden)^{2/3}\equiv\eps(n_{\ua},\ka),
\label{eq:nor}
\ee
where the dimensionless parameter $\xi_{\rmn}(\ka)$ accounts for the interactions, and $m_{\rm \ka}=(\mup\mdown)/(\mup+\mdown)$ is the reduced mass. First results based on quantum Monte Carlo (QMC) calculations suggest that the dependence of the interaction parameter $\xi_{\rmn}(\ka)$ on the mass ratio $\ka$ is very weak \cite{note1} so that we can set $\xi_{\rmn}(\ka)\equiv\xi_{\rmn}=0.56$ \cite{Carlsonxi,Stefanoxi} also for $\ka\neq1$. Therefore the effect of unequal masses on the energy of the unpolarized normal state enters only through the reduced mass $m_{\ka}$. We can define $B$  as a function of the mass ratio by requiring that the energy of the normal state be reproduced by Eq.(\ref{eq:eoslobo}) for a concentration $x=1$ with the $\ka$-dependent interaction parameters $A$ and ${m^*}/{m_{\da}}$ given in \cite{Combescot}. Then the generalization of Eq.(\ref{eq:eoslobo}) to the unequal mass case yields
\beq
\frac{E(x,\kappa)}{N_\ua}&=&\frac{3}{5}\eup\left(1-A(\kappa)
x+\frac{F(\kappa)^{-1}}{\kappa}x^{5/3}+B(\kappa) x^2\right)\nonumber\\
&=&\frac{3}{5}\eup g(x,\kappa)\equiv\eps_{\rmn}(x,\ka).  
\label{eq:enkappa}
\enq

\section{Phase Diagram}
\label{pd}

In terms of the mass ratio $\ka$ the superfluid energy takes the form
\be
\frac{E_{\rm S}(\ka)}{N_{\rm S}}= \xi_{\rms}(\ka)\frac{3}{5} \frac{\hbar^2}{4m_{\rm \ka}} (6 \pi^2 n_{\rm S})^{2/3}\equiv\eps_{\rms}(n_{\rms},\ka),
\label{eq:sur}
\ee
where $N_{\rm S}$ is the number of atoms in the superfluid phase, $n_{\rm S}$ the superfluid density, $m_{\ka}$ the reduced mass, and $\xi_{\rms}(\ka)$ accounts for the interactions in the superfluid. 
Also in the superfluid phase the coefficient $\xi_{\rms}(\ka)$ has only a very weak dependence on the mass ratio \cite{note1} so that we can set $\xi_{\rms}(\ka)\equiv\xi_{\rms}=0.42$ \cite{Carlsonxi,Stefanoxi} as in the equal mass case. 

In order to establish the phase diagram for the system we address the equilibrium conditions for the phase separation of the superfluid and  normal state in the bulk.
We start by writing down the energy of the system at zero temperature
\beq
E&=&2 \int d{\bf r}\ \left\lbrack \frac{}{}\epsilon_{\rms}(n_{\rms},\ka)n_{\rms}-\mu_{\rms}^0n_{\rms}\frac{}{}\right\rbrack \nonumber\\
& & +\int d{\bf r}\left\lbrack\frac{}{}\eps_{\rmn}(x,\ka)n_{\ua}-\mu_{\ua}^0n_{\ua}-\mu_{\da}^0n_{\da}\frac{}{}\right\rbrack,
\label{eq:energybulk}
\enq
where $\eps_{\rms}(n_{\rms},\ka)$ and $\eps_{\rmn}(x,\ka)$ are the energy densities per particle, $n_{\rms},n_\ua$ and $n_\da$ the densities, $\mu^0_{\ua}$ and $\mu^0_\da$ the chemical potentials of the spin-$\ua$ and spin-$\da$ component, respectively, and $\mu^0_{\rms}=(\mu_{\ua}^0+\mu_{\da}^0)/2$ is the superfluid chemical potential. 

To find the equilibrium conditions we minimize the energy with respect to the densities of the superfluid and normal phase, and we find the chemical potentials
\be
\mu_{\rm S}^0=\xi_{\rms}\frac{\hbar^2}{4m_{\rm \ka}}(6 \pi^2 n_{\rm S})^{2/3},
\label{eq:musbulk}
\ee
\be
\mu_\ua^0=\left(g(x,\ka)-\frac{3}{5}x g'(x,\ka)\right)\frac{\hbar^2}{2\mup}(6 \pi^2 n_\ua)^{2/3},
\label{eq:mupbulk}
\ee
\be
\mu_\da^0=\frac{3}{5}g'(x,\ka)\frac{\hbar^2}{2\mup}(6 \pi^2 n_\ua)^{2/3},
\label{eq:mdbulk}
\ee
where prime means the derivative with respect to $x$.
Eventually requiring that the pressure of the two phases be the same yields
\be
\left(n_{\rms}^2\frac{\partial{\epsilon_{\rms}}}{\partial n_{\rms}}\right)=\frac{1}{2}\left(n_{\ua}^2\frac{\partial{\epsilon_{\rmn} (x,\ka)}}{\partial n_{\ua}}+n_{\da}n_{\ua}\frac{\partial{\epsilon_{\rmn}(x,\ka)}}{\partial n_{\da}}\right).
\label{eq:eqpress}
\ee
\begin{figure}[htb]
	\centering
		\includegraphics[height=5cm]{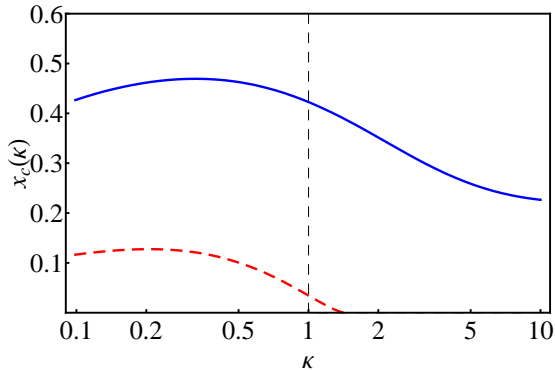}
	\caption{(Color online) Critical concentration $x_c(\ka)$ for the bulk system to phase separate as a function of the mass ratio $\ka$ (solid blue). In comparison, also the concentration derived from the BCS mean-field solutions at unitarity is shown (dashed red).}
	\label{fig:conlo}
\end{figure}
Making use of Eqs.(\ref{eq:enkappa}) and (\ref{eq:sur}) we can write the equal pressure condition as the density jump
\be
\frac{n_{\ua}(x,\ka)}{n_{\rms}(\ka)}=\left(\frac{(1+\frac{1}{\ka})\xi_{\rms}}{g(x,\ka)}\right)^{3/5}.
\label{eq:jp}
\ee
From Eqs.(\ref{eq:musbulk})-(\ref{eq:eqpress}) we obtain the implicit equation 
\beq
g(x(\ka))&+&\frac{3}{5}[1-x(\ka)]g'(x(\ka))-\nonumber\\
&-& \left\lbrack\left(1+\frac{1}{\ka}\right)\xi_{\rms}\right\rbrack^{3/5}[g(x(\ka))]^{2/5}=0,
\label{eq:condv}
\enq
which gives the Chandrasekhar-Clogston limit $\xc$ in dependence on the mass ratio $\ka$.
For values smaller than $x_{\rm c}(\ka)$ the system remains normal, while for $x>x_{\rm c}(\ka)$ the system starts nucleating the superfluid and phase separates into those two states. 
In Fig. \ref{fig:conlo} we plot $x_c(\ka)$ for mass ratios $0.1<\ka<10$ (blue solid line). Comparing with $\xc(\ka=1)=0.42$ we find that for mass ratios $\ka>1$ the concentration needed to create a superfluid phase decreases, while for mass ratios $\ka<1$ it first increases and reaches a maximum value at $\ka\sim0.3$.
In the same figure we plot $\xc(\ka)$ as resulting from the BCS mean-field approach at unitarity (red dashed line, see also  e.g. \cite{yip}). In the latter treatment interactions in the normal phase are not taken into account, and hence its energy  is just the sum of the $\ua$ and $\da$ components, $E_{\rm N}^{\rm{BCS}}=E_{{\rm F}\ua}N_{\ua}+E_{{\rm F}\da}N_{\da}$, and the interaction parameter for the superfluid is $\xi_{\rm S}^{\rm {BCS}}=0.59$. The  significant quantitative difference between the two curves proves the importance of interactions \cite{alessio}.

It is worth noticing that $\xc$ is sensitive to the actual value of the parameters used in Eq.(\ref{eq:enkappa}). 
Since an exact calculation of the parameter $B(\ka)$ in the case of unequal masses is still lacking, the interpolated value of $B(\ka)$ might be a significant source of error. 
As for $\ka>>1$ the kinetic energy of the quasiparticles becomes irrelevant (see Eq.(\ref{eq:enkappa})), the Chandrasekhar-Clogston limit is only determined by the values of $A(\ka)$ and $B(\ka)$. Thus an uncertainty in $B$ affects more our predictions. On the other hand  for $\ka>>1$ a polarized superfluid phase might have to be included in the description of the system so that the two-phase assumption will not longer be valid  (see also discussion in Sec. \ref{con}).

We varied the value of $B(\ka)$ by $\pm 10\%$ to see its final impact on the value of $\xc(\ka)$, and we find that the variation in $\xc(\ka)$ is around $\mp5\%$ for mass ratios $\ka\leq1$, while for $\ka>1$ it is $\mp 10\%$.

In terms of the chemical potentials of the $\ua$ and $\da$ components the phase transition is characterized by the critical value $\etc(\ka)=({\mu_\da}/{\mu_\ua})_{\xc}$. From the knowledge of $\etc(\ka)$ we are able to determine the coexistence lines between the superfluid and the normal phase.

We represent the different homogeneous phases employing the $\mu-h$ phase diagram, where $2\mu=\mu_\ua+\mu_\da$ and $2 h=\mu_\ua-\mu_\da$.
The transition line between the superfluid ($\rms$) and partially polarized ($\rm PP$) phase is given by
\beq
\mu_{\rm S}^{h>0}&=&\ \frac{1+\etc(\ka)}{1-\etc(\ka)}h,\nonumber\\
\mu_{\rm S}^{h<0}&=&-\frac{1+\etc(\frac{1}{\ka})}{1-\etc(\frac{1}{\ka})}h,
\label{eq:musf}
\enq
and stands for the first-order phase transition between the unpolarized superfluid and the partially polarized normal phase.

The second-order phase transition between the partially polarized ($\rm PP$) and the fully polarized ($\rm FP$) phase occurs at $x=0$, which corresponds to ${\mu_\da}/{\mu_\ua}=-{3}/{5}A(\ka)$, and thus the coexistence line is given by
\beq
\mu_{\rm PP}^{h>0}&=&\ \frac{1-\frac{3}{5}A(\ka)}{1+\frac{3}{5}A(\ka)}h,\nonumber\\
\mu_{\rm PP}^{h<0}&=&-\frac{1-\frac{3}{5}A(\frac{1}{\ka})}{1+\frac{3}{5}A(\frac{1}{\ka})}h.
\label{eq:mun}
\enq

Finally, the transition line between the fully polarized gas and the vacuum is given by the simple $\ka$-independent relation
\be
\mu_{\rm FP}=-\mid h\mid.
\label{eq:mup}
\ee

\begin{figure}[htb]
	\centering
		\includegraphics[height=5cm]{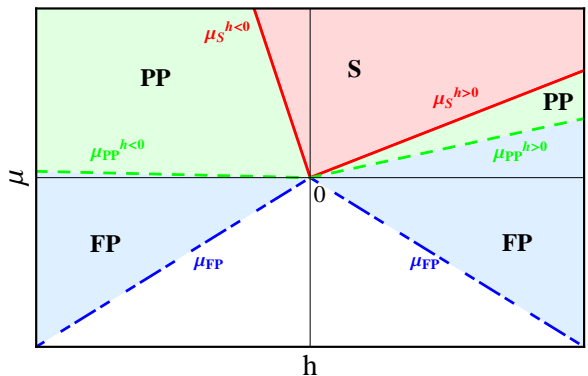}
	\caption{(Color online) For $\ka=2.2$ the phase diagram is asymmetric. Shown are the superfluid $\rm S$ (solid red lines), partially polarized $\rm PP$ (dashed green) and fully polarized $\rm FP$ (dot-dashed blue) regions.}
	\label{fig:Bpd}
\end{figure}

The phase diagram for unequal masses is not symmetric with respect to zero effective magnetic field $h$ as can be seen in Fig. \ref{fig:Bpd}, where we choose $\ka=2.2$ corresponding to the case of a $^{87}$Sr -$^{40}$K mixture \cite{felikx}. While the superfluid $\rms$ moves clockwise (anticlockwise) for $\ka>1$ ($\ka<1$), the partially polarized $\rm {PP}$ moves in the opposite direction, see e.g. Figs. \ref{fig:Apd} and \ref{fig:Bpd}. {In all the figures we use a solid line for the first-order phase transition, a dashed line for the second order phase transition, and a short-dashed-long-dashed line for the transition to the vacuum.}

Such an asymmetry in phase diagrams is general for this system and has been already noticed by Parish {\sl {et al.}} \cite{lama} in the $T/\mu$ vs $h/\mu$ phase diagram, and by Iskin  and S\'a de Melo \cite{melo} in the $P=(N_{\ua}-N_{\da})/(N_{\ua}+N_{\da})$  vs $(1/k_{F,+},a_{F})$ diagram.

In particular, from Eq.(\ref{eq:musf}) we can identify a critical mass ratio $\ka^*=2.73$ above which the superfluid region has moved entirely to the $h>0$ plane (see e.g. Fig. \ref{fig:66}(a)). 
This shift of the superfluid region above a certain mass ratio $\ka$ has also been identified by Parish {\sl et al.} \cite{lama} applying BCS mean-field theory yielding $\ka^*\sim3.95$. 

At the same time, for $\ka>\ka^*$ the sum of the spin-$\ua$ and spin-$\da$ densities in the partially polarized phase is bigger than the superfluid density, $(n_{\ua}+n_{\da})>2n_{\rms}$. This anticipates the fact that in a trap the heavy partially polarized phase can sink towards the center, while the superfluid will form a spherical shell around it even if the two species feel the same trapping potential. 
This peculiar formation of a \lq\lq sandwiched\rq\rq superfluid  has been previously identified also in \cite{yip,duan,lama}.

\section{Trapped Gas}
\label{trap}

Having constituents with different masses and hence different magnetic and optical properties permits to engineer different configurations in the trap depending on the mass ratio, the polarization, and the choice of the trap parameters.

In order to study the trapped case we assume that the external potential is harmonic of the form $V_{\sigma}({\bf r})=\frac{1}{2}\alpha_{\sigma}r^2$ where $\alpha_\sigma=m_{\sigma}\om^2_{\sigma}$ with $\sigma=\ua,\da$, and that the local density approximation is applicable. Thus the configuration in the trap is found by using the expression $\mu_{\sigma}=\mu_{\sigma}^0-\frac{1}{2}\as r^2$  leading to
\beq
\mu&=&\mu^0_{\ua}\left[\frac{1+\eta_0}{2}-\frac{1}{2}\left(1+\frac{\ad}{\au}\right)\frac{r^2}{(R^0_{\ua})^2}\right],\nonumber\\
h&=&\mu^0_{\ua}\left[\frac{1-\eta_0}{2}-\frac{1}{2}\left(1-\frac{\ad}{\au}\right)\frac{r^2}{(R^0_{\ua})^2}\right],
\label{eq:ldagenk}
\enq
where we define $\eta_0=\mda^0/\mua^0$ as the central imbalance of the system, and $(R^0_{\ua})^2=2\mu^0_{\ua}/\au$. 
Note that if $\au=\ad$ the effective magnetic field $h$ does not depend on the position in the trap but is only a function of the central imbalance $\eta_0$.
Concerning the central imbalance of the chemical potentials we have that if $\eta_0<\eta_{\rm c}(\ka)$, there is no superfluid and the system consists only of the partially and fully polarized component. In the case that $\eta_0>\eta_{\rm c}(\ka)$, we have a superfluid component whose fraction is determined by the value of $\eta_0$.

In the following we will describe three different cases with different values of the polarization
\be
P=\frac{N_{\ua}-N_{\da}}{N_{\ua}+N_{\da}},
\ee
where the interplay between the asymmetry in the masses and in the trapping potential gives rise to configurations.

\subsection{Unequal masses with equal trapping}
\label{uet}

\begin{figure}[htb]
	\centering
		\includegraphics[height=10cm]{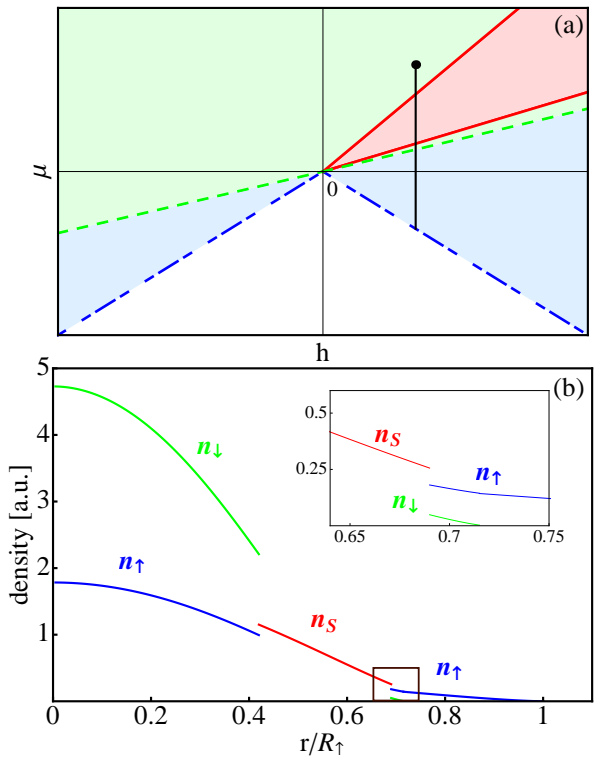}
	\caption{(Color online) (a) Phase diagram for $\ka=6.7$, corresponding to a $^{40}$K-$^{6}$Li mixture, and the LDA line (black vertical) for a central imbalance $\eta_0=0.3$ (black dot) and $\au=\ad$. (b) Density profiles for $P=-0.13$; the inset shows a zoom into the outer superfluid-"light" normal border.}
	\label{fig:66}
\end{figure}

We first analyze the situation when the spin-$\ua$ and spin-$\da$ components have different masses $\ka\neq1$ but feel the same restoring forces $\au=\ad$.
This would be the case, for example, if the fermions are trapped magnetically and have identical magnetic moments.
For equal populations and for mass ratios in the range $0.36<\ka<2.73$ the system is completely superfluid.
In the opposite case, the system can never be completely superfluid even if the populations are equal. Therefore, we can also have the particular configuration of a system consisting only of a partially polarized phase (without any fully polarized part).

If $\eta_0>1/\etc({1}/{\ka})$ the trapped system will consist of a  three-shell configuration, where the superfluid is sandwiched between a \lq\lq heavy\rq\rq normal phase (heavy spin-$\da$ are the majority) at the center of the trap, and a \lq\lq light\rq\rq normal phase (light spin-$\ua$ are the majority) in the outer trap region.

As an example we choose the mass ratio $\ka=6.7$ corresponding to a $^{40}$K -$^{6}$Li mixture \cite{haensch,felikx}.
The phase diagram of the system is shown in Fig. \ref{fig:66}(a) together with the LDA line for a central imbalance $\eta_0=0.3$ (black dot).
The intersection of the LDA line with the coexistence lines determines the radii of the configuration, from which we are able to calculate the density profiles. These are shown in Fig. \ref{fig:66}(b) for a polarization $P=-0.13$.

The density jump (or drop) between the superfluid and both normal phases is a function of $\ka$ according to Eq.(\ref{eq:jp}). For $\ka=6.7$ at the \lq\lq heavy\rq\rq normal - superfluid border, $n_{\da}\sim1.92n_{\rms}$ and $n_{\ua}=\xc(\frac{1}{\ka})n_{\da}\sim0.86n_{\rms}$, while at the superfluid-\lq\lq light\rq\rq normal border $n_{\ua}\sim0.71n_{\rms}$ and $n_{\da}=\xc(\ka)n_{\ua}\sim0.17n_{\rms}$ (see inset Fig. \ref{fig:66}(b)). Note that this is quite different compared to the equal mass case, where the jump between the superfluid and the majority component is $n_{\ua}\sim1.01n_{\rms}$ and hence the spin-$\ua$ density is practically continuous.

\subsection{Unequal masses with trapping anisotropy} 

\begin{figure}[htb]
	\centering
		\includegraphics[height=10cm]{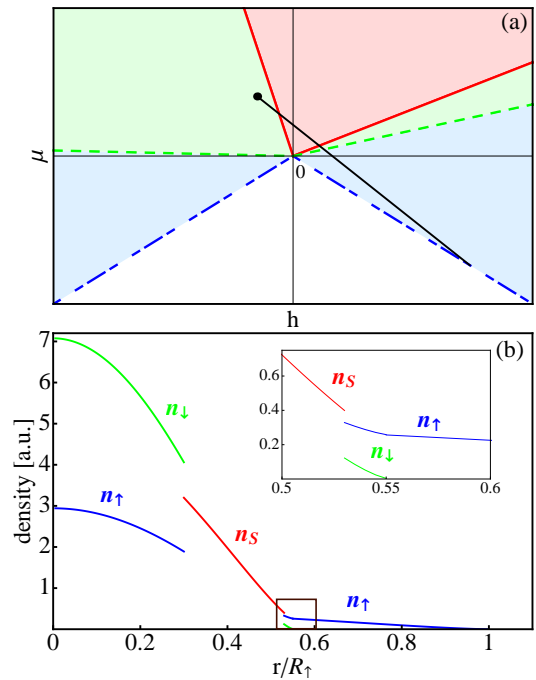}
	\caption{(Color online) (a) Phase diagram for $\ka=2.2$ and LDA line  for $\eta_0=2.1$ (black dot) and $\ad/\au=8$. With this choice, it crosses the "heavy normal", superfluid, "light normal" and fully polarized phases. (b) Density profiles for a global polarization $P=0$; the inset shows a zoom into the superfluid-"light" normal border.}
	\label{fig:diffapd}
\end{figure}

Using unequal restoring forces for the trapped atoms the mass ratio for having a sandwiched superfluid needs not to be necessarily bigger than the critical value $\ka^*$.  
In order to have a three-shell configuration the condition is
\beq
\mu<\mu_{\rm S}^{h<0} \Rightarrow \frac{{\ad}}{{\au}}=\ka\frac{\om^2_{\da}}{\om^2_{\ua}}>\frac{1}{\etc(\frac{1}{\ka})}.
\label{eq:trapcon}
\enq 
For example, for equal trapping frequencies, Eq.(\ref{eq:trapcon}) simplifies to $\ka>[{1/\etc({1}/{\ka})}]$ resulting in the critical mass ratio $\ka_{\rm c}\sim1.95$, while for equal oscillator lengths one gets $\ka_{\rm c}\sim6.7$.
In Fig. \ref{fig:diffapd}(a) we show the $\mu-h$ phase diagram of such a particular configuration, where we choose $\ka=2.2$ corresponding to a $^{87}$Sr -$^{40}$K mixture \cite{felikx}. The LDA line is drawn for the values  $\eta_0=2.1$ (black dot) and $\ad/\au=8$.

In the density profiles as shown in Fig. \ref{fig:diffapd}(b) we have chosen the parameters such that the resulting global polarization is $P=0$.

\subsection{No trapping for $\ua$ component}

An interesting limiting case is when one of the elastic constants $\alpha_{\sigma}$ is zero (or very small), implying that one of the components would not be confined in absence of interspecies atomic forces.

\begin{figure}[htb]
\centering
		\includegraphics[height=10cm]{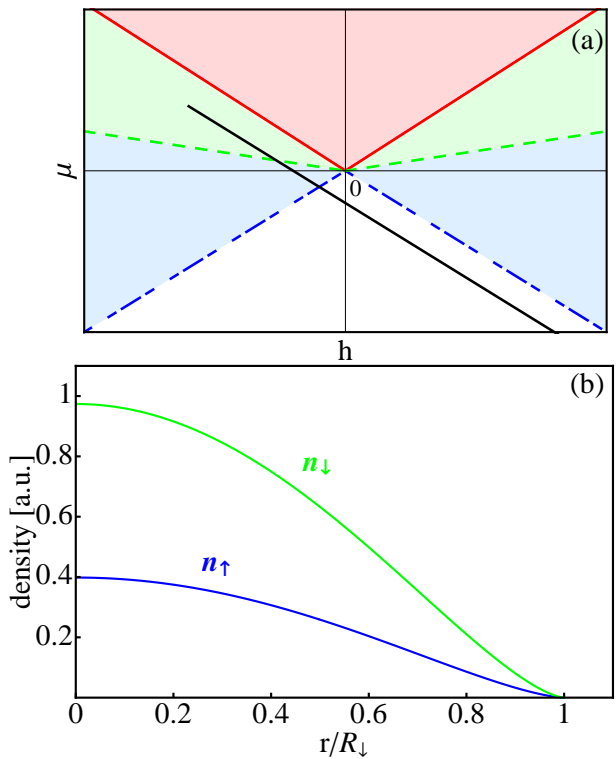}
	\caption{(Color online) (a) Phase diagram for $\ka=1$. The LDA line represents the case $\au=0$. (b) Density profile in the limiting case $\au=0$ and $P=-0.42$.}
	\label{fig:ntpd}
\end{figure}

If we assume that $\au\rightarrow0$, the LDA line in the $\mu$-$h$ phase diagram is parallel to the polarized-vacuum transition line as shown in Fig. \ref{fig:ntpd}(a).

Let us start considering the equal mass, $m_\ua=m_\da=m$, highly unbalanced $N_{\da}\gg N_{\ua}$ case. The densities are easily found to be \cite{alessio}
\beq
\mu_{\da}^0&=&\frac{\hbar^2}{2m}\left[6\pi^2n_{\da}({\bf r})\right]^{2/3}+V_{\da}({\bf r}),\nonumber\\
\mu_{\ua}^{0'}&=&\frac{\hbar^2}{2m^*}\left[6\pi^2n_{\ua}({\bf r})\right]^{2/3}+V'_{\ua}({\bf r}),
\enq
where $\mu_{\ua}^{0'}=\mu_{\ua}^{0}+\frac{3}{5}A\mu_{\da}^0$, $V'_{\ua}({\bf r})=V_{\ua}({\bf r})+\frac{3}{5}AV_{\da}({\bf r})$ and $A\equiv A(\ka=1)$.
From these equations it is clear that if $V_\ua\rightarrow 0$, the $\ua$-atoms feel nevertheless the renormalized potential $\frac{3}{5}AV_{\da}({\bf r})$ and are confined due to the interaction with the $\da$-component. In this regime $\mu_\ua^0$ is negative and in the limit of a single $\ua$-atom, i.e. $\mu_\ua^{0'}\rightarrow 0$, it takes the value $\mu_\ua^0=-3/5 A \mu_\da^0$, corresponding to a polarization $P=-1$.
This induced trapping mechanism would not be predicted by a BCS mean-field description, where interactions are absent in the normal phase, and the $\ua$-atoms cannot be confined by the $\da$-atoms. 

Increasing the number of $\ua$ particles, the LDA line moves upward until it crosses the origin of the phase diagram, corresponding to $\mu_\ua^0=0$, and the system remains normal since for equal masses the slope of the superfluid-partially polarized coexistence line is bigger than the slope of the LDA line, i.e. $\eta_c(1/\ka=1)>0$ (see Eq.(\ref{eq:musf})). Moreover, in this case the $\da$-fully polarized phase is absent as the radii of the $\da$ and $\ua$ species coincide, and in this limit the polarization approaches the value $P=-0.42$.

If we further increase $N_\ua$ we enter in a three-shell configuration including an intermediate superfluid component. But since in this case the atoms of species $\ua$ are no longer confined, they escape from the trap, and the system goes back to the normal state previously described. Hence we can never find a stable configuration containing a superfluid region, and the polarization of the system will always be in the range $-1<P\le-0.42$.

Note that the same scenario is valid for $m_\da>0.9m_\ua$, where $\eta_c(1/\ka)$ is positive and the range of the polarization is between $P=-1$ and an upper value which is dependent on $\ka$. 

Interestingly, in the case $m_\da<0.9m_\ua$, for which $\eta_c(1/\ka)<0$, we find that adding $\ua$-atoms we end up in a superfluid state \cite{orso}
characterized by a density profile given by 
\be
\mu_{\rms}^0=\xi_{\rms}\frac{\hbar}{2m}\left[6\pi^2n_{\rms}({\bf r})\right]^{2/3}+\widetilde{V}({\bf r}), 
\ee
where $\widetilde{V}({\bf r})=\frac{1}{4}m\om_{\da}^2r^2$ is the effective potential felt by the superfluid. This configuration would correspond to a LDA line which stays entirely in the superfluid region, crossing the origin of the phase diagram.
The value of the polarization for $m_\da<0.9m_\ua$ covers the entire range  $-1<P\le0$.

From the experimental point of view the above configurations could be in principle reached starting with both the trapping frequencies different from zero and a certain initial polarization, and then opening adiabatically the trap for the $\ua$-atoms. For instance, starting with only a superfluid in the trap the final state of the system will be simply a superfluid with a bigger radius for $m_\da<0.9 m_\ua$, while it will be a normal state in which both components have the same radius (see Fig. \ref{fig:ntpd}(b)) if $m_\da>0.9m_\ua$.
  
\section{Conclusions}
\label{con}

We have studied the zero temperature $\mu$-$h$ phase diagram of the unitary Fermi gas in the case of unequal masses, assuming phase separation between an unpolarized superfluid and a polarized normal phase. The latter is described by an equation of state which, unlike in the BCS mean-field treatment, takes into account the effect of the strong interaction. 
As we have shown, this has a dramatic impact on the results such as the Chandrasekhar-Clogston limit needed to start nucleating a superfluid. 

Using LDA we have determined how the trapped configuration depends on the trapping potential, the mass ratio, and the polarization.
Many different configurations are possible. Among them it is worth mentioning the three-shell configuration \cite{yip,duan,lama}, where the superfluid is sandwiched between a \lq\lq heavy\rq\rq normal phase at the center and a \lq\lq light\rq\rq normal phase towards the edges of the trap. Note that the shells can occupy quite small regions, and we cannot exclude that surface tension plays an important role in this case.

We can also have non-trivial configurations even if one of the two components is not trapped, but still remains confined due to the interaction induced trapping. Such configurations can be experimentally obtained by adiabatically opening the trap for one of the two species.

An important issue is the existence of other phases at unitarity. In the present work we assume that only two phases are possible, and hence we have not considered any polarized superfluid state. For the equal mass case the assumption seems to be correct and is theoretically understood by comparing the phase separated state energy with the polarized superfluid energy calculated via Monte Carlo, as in e.g. \cite{polarizedSS}. The same information is not yet available for the unequal mass case. However, taking the quasi-particle point of view in \cite{polarizedSS} and the recent calculation for equal population by Baranov {\sl et al.} \cite{bara}, it seems that when the mass of the minority component is much bigger than the one of the  majority component, the polarized superfluid phase should be included in the description, as predicted by  mean-field theory. Theoretical work in this direction is in progress.

{{\sl{Note added.}}  Recently, a Monte Carlo calculation for the system considered here for $\ka=6.5$ was posted in \cite{Gez}. Although an accurate comparison has not yet been done, it seems that the Monte Carlo analysis is in agreement with our description.}

\section{Acknowledgements}
We thank Carlos Lobo, Stefano Giorgini, and Rudolf Grimm for useful discussions. We acknowledge support by the Ministero dell'Istruzione, dell'Universit\`a e della Ricerca (MIUR) and by the EuroQUAM FerMix program.

\end{document}